\documentclass[a4paper,fleqn,usenatbib]{mnras}

\usepackage{newtxtext,newtxmath}

\usepackage[T1]{fontenc}
\usepackage{ae,aecompl}
\usepackage{soul}

\usepackage{graphicx}	
\usepackage{amsmath}	
\usepackage{amssymb}	
\usepackage{bm}		
\usepackage{hyperref}
\usepackage{subfigure}
\usepackage{romannum}

\newcommand{\diff}{{\rm d}}

\title[Probing iPTF16axa]{Probing the tidal disruption event iPTF16axa with CLOUDY and disc-wind models}

\author[Mageshwaran et al.]{
T. Mageshwaran$^{1}$\thanks{E-mail: tmageshwaran2013@gmail.com}, Gargi Shaw$^{2}$ and Sudip Bhattacharyya$^{2}$
\\
$^{1}$Department of Space Science and Astronomy, Chungbuk National University, 12 Gaeshin-dong, Heungduk-gu, Cheongju 361-763, Korea \\
$^{2}$Department of Astronomy and Astrophysics, Tata Institute of Fundamental Research, 1 Homi Bhabha Road, Mumbai 400005, India
}

\date{Accepted 2022 November 25. Received 2022 November 25; in original form 2022 February 16}

\pubyear{2022}

\begin{document}
\label{firstpage}
\pagerange{\pageref{firstpage}--\pageref{lastpage}}
\maketitle


\begin{abstract}
We present both a disc-wind model on the optical/UV emission continuum and CLOUDY modelling on the spectral lines of the tidal disruption event (TDE) iPTF16axa to understand the disc-wind emission and the properties of the atmosphere that impacts the line luminosity of the TDE. Assuming the optical/UV emission from the wind due to the disc super-Eddington phase, we use the steady structured disc-wind model with a spherical wind with constant velocity to fit the observations on multiple days. The extracted parameters are stellar-mass $M_{\star} = 6.20 \pm 1.19 M_{\odot}$, disc radiative efficiency $\log_{\rm 10}(\eta) = -1.22 \pm 1.327$, wind inner radius $r_l = (2.013 \pm 0.551) \times 10^{14}~{\rm cm}$ and velocity $v_w = 18999.4 \pm 1785.1 ~{\rm km~s^{-1}}$. The photosphere temperature for wind emission is $ \sim 2 \times 10^4~{\rm K}$ and the disc single blackbody temperature is $\sim 0.995 \times 10^5~{\rm K}$. We also perform CLOUDY modelling to explain the observed He and H line luminosities that estimate a wind inner radius $r_l = 7.07 \times 10^{14}~{\rm cm}$ and velocity $v_w = 1.3 \times 10^4~{\rm km~s^{-1}}$. The independent analyses of iPTF16axa using CLOUDY and disc-wind models show comparable results that agree with observations. The CLOUDY modelling finds that both the super solar abundance of He and a smaller He II line optical depth is responsible for the enhancement of He II line luminosity over the H$\alpha$ line luminosity. The super-solar abundance of He II agrees with a relatively large stellar mass and suggests that the disrupted star might have been a red giant.     
\end{abstract}




\section{Introduction}

A star orbiting around black holes is tidally disrupted when the tidal force exceeds the self-gravity of the star \citep{1976MNRAS.176..633F,1988Natur.333..523R}. The fraction of stellar mass stripped during disruption depends on the orbital pericenter and stellar density profile \citep{2013ApJ...767...25G}. The tidal radius below which the star is fully disrupted occurs is $r_t= (M_{\bullet}/M_{\star})^{1/3} R_{\star}$, where $M_{\bullet}$ is the black hole mass, and $M_{\star}$ and $R_{\star}$ are the stellar mass and radius respectively \citep{1976MNRAS.176..633F}. The disrupted debris following an orbit returns to the pericenter with a mass fallback rate that at late time $t$ follows $t^{-5/3}$ evolution \citep{2009MNRAS.392..332L,2015ApJ...814..141M}. The interactions between the debris streams result in an exchange of angular momentum and circularization leading to the formation of an accretion disc \citep{2013MNRAS.434..909H,2016MNRAS.455.2253B}. The thermal luminosity from TDEs is observed over a wide spectral range from infrared, optical, UV, and X-rays. The relativistic jet produces non-thermal emissions in X-ray, gamma-ray, and radio bands.

TDE detection has increased in a few decades with multi-wavelength observations from space and ground-based telescopes. The TDEs are observed in multi-wavelength bands such as in X-rays by XMM-Newton \citep{2007A&A...462L..49E,2014A&A...572A...1S} and Chandra X-ray Observatory \citep{2010ApJ...722.1035M,2013MNRAS.435.1904M}, in optical bands \citep{2012Natur.485..217G, 2014ApJ...793...38A, 2014ApJ...780...44C, 2014MNRAS.445.3263H} and in UV \citep{2008ApJ...676..944G, 2009ApJ...698.1367G}. The observations of TDE candidates in multi-wavelength bands and their high-resolution spectra provide a deeper understanding of the accretion discs and their evolutions. 

A transient is classified as a TDE based on the late time decline of the bolometric luminosity (close to $t^{-5/3}$), and the broad emission lines with blue continuum \citep{2016MNRAS.463.3813H,2017ApJ...844...46B,2017ApJ...842...29H}. The TDE observations initially were modelled with the luminosity $L \propto (t-t_D)^{-5/3}$, where $t_D$ is the disruption time  \citep{2008ApJ...676..944G,2011ApJ...741...73V}, whereas \citet{2014MNRAS.445.3263H} [ASASSN-14li] and \citet{2016MNRAS.463.3813H} [ASASSN-15oi] showed that the luminosity is best fitted with the exponential decline. The evolution of luminosity in various spectral bands depends on the disc evolution in sub-Eddington and super-Eddington phases. For black hole mass $M_{\bullet} \lesssim ~{\rm few}~10^{7} M_{\odot}$, the mass fallback rate is greater than the Eddington rate $\dot{M}_E$ for a period of weeks to years depending on the black hole mass; here $ \dot{M}_E = L_E/(\eta c^2)$, where $L_E$ is the Eddington luminosity, $c$ is the speed of light and $\eta$ is the radiative efficiency. 

The disc dominated by radiation pressure and viscous heating flux equal to the radiative flux is thermally unstable \citep{1974ApJ...187L...1L}. The disc near the super-Eddington mass accretion rate is dominated by the radiation pressure and the advective energy flux is crucial. The strong radiation force at the disc surface in the super-Eddington phase results in an outflow \citep{2020SSRv..216..114R}. 
In most of the TDEs, the optical/UV emission dominates at early times and the X-ray flux is lower than that expected from the disc. The temperature obtained using a blackbody model fit to the X-ray spectrum of the source such as XMMSL1 J061927.1-655311 is $\sim 1.4 \times 10^6$ K \citep{2014A&A...572A...1S}, ASAS-SN 14li is $\sim 10^5$ K \citep{2016MNRAS.455.2918H}, Abell-1795 is $\sim 1.2 \times 10^6$ K \citep{2013MNRAS.435.1904M} and NGC-3599 is $\sim 1.1 \times 10^6$ K \citep{2007A&A...462L..49E}, whereas the temperature from the optical/UV observations is $\sim 10^4$ K \citep{2020SSRv..216..114R}. The X-ray emission from the disc is reprocessed to the lower optical/UV wavelengths by the reprocessing layer which may be due to the outflowing wind \citep{2009MNRAS.400.2070S,2020ApJ...894....2P}, or the material ejected during the interaction of debris stream near pericenter \citep{2014ApJ...783...23G}. The reprocessing and the emission from the outflowing wind is a complex mechanism, requiring in detail the radiative transfer model.  

However, \citet{2009MNRAS.400.2070S} [hereafter SQ2009] constructed the steady structure slim disc accretion model with mass accretion rate ($\dot{M}_a$) that follows the mass fallback rate ($\dot{M}_{\rm fb}$). They assumed the outflowing wind to be spherical with constant velocity and the mass outflow rate $\dot{M}_{\rm out} = f_0 \dot{M}_a$, where $f_0$ is taken to be a constant. \citet{2015ApJ...814..141M} [hereafter MM2015] included the orbital energy and angular momentum of the star prior to disruption in calculating the mass fallback rate and used the $f_0$ that is a function of mass accretion rate to calculate the emission from the slim disc and the wind. \citet{2014ApJ...781...82C} developed a steady structured Zero Bernoulli Accretion (ZEBRA) Flow for a hyperaccreting disc ($\dot{M}_a/\dot{M}_E \gg 1$), where the highly critical accretion results in a bipolar jet from the inner radius. The photons interact with the electrons through Thomson scattering results in a momentum transfer causing a radial inflow of the matter and is a source of viscosity. For an isotropic distribution of photons, \citet{1968ApJ...151..431M} derived the viscous stress which is the radiative viscosity and is used by \citet{1992ApJ...384..115L}. A time-dependent and power-law self-similar solution for an accretion disc with outflows and radiative viscosity was constructed by \citet{2021NewA...8301491M}. 

The outflowing wind is modelled with spherical geometry and power law density given by $\rho(r) \propto r^{-2}$ \citep{2009MNRAS.400.2070S}. The photons are coupled with the gas until the trapping radius ($r_{\rm tr}$) from where the diffusion time is higher than the dynamical time. The wind expands adiabatically for $r < r_{\rm tr}$ and evolves via the diffusion model for $r > r_{\rm tr}$ until it is radiated. \citet{2020ApJ...894....2P} used the effective opacity $\kappa_{\rm eff} \simeq \sqrt{3 \kappa_s \kappa_a}$ in the diffusion regime, where $\kappa_s$ and $\kappa_a$ are Thomson and Kramer opacities. They showed that the blackbody radius corresponding to the observed temperature is much less than the actual colour radius (optical depth $\tau = 1$). Thus, they inferred that the models that are using the blackbody emission for fitting are not fitting for the true emission radius. These models assumed the bolometric luminosity from the central source is $L \propto \dot{M}_{\rm fb} c^2$. \citet{2020ApJ...905L...5U} developed a similar model and included the recombination temperature to fit the observed bolometric luminosity and temperature to extract the mass outflow rate and colour radius. We use the slim disc model as the source of central emission with a spherical wind model to fit the luminosity in various spectral bands.

\citet{2016ApJ...827....3R} considered a static atmosphere with fixed inner radii, envelope mass $M_{\rm env}$ and $\rho(r) \propto r^{-2}$ in their radiative transfer code with a given luminosity $L$ to extract the reprocessed spectrum. The scattering and absorption of photons in the atmosphere increase with the optical depth that depends on the density and temperature of the medium, and the photon frequency. They have shown that the dominance of helium lines over hydrogen is because of the suppression of Balmer lines due to optical depth effects. However, the atmosphere evolves where density changes with time which impacts the photosphere radius and the temperature. The hydrodynamical simulation of \citet{2014ApJ...783...23G} found that the structure and evolution of the envelope (reprocessing layer) depends on the details of how energy is injected by the infalling debris and may exist even if $\dot{M}_a < \dot{M}_E$. They calculated the reprocessing of luminosity $L$ using an empirical formulation for photosphere radius and temperature in terms of mass fallback rate. We use the disc-wind model to fit the continuum spectrum obtained using optical/UV photometry to extract the atmosphere details, the temperature, and the photosphere radius and constraint them using the atmosphere details inferred from CLOUDY modelling to the observed spectral line luminosities.  

The photoionization code CLOUDY \citep{2013RMxAA..49..137F,2017RMxAA..53..385F} has been used to explain the dominance of the helium line over the Balmer lines for the TDE PS1-10jh \citep{2012Natur.485..217G,2015MNRAS.454.2321S}. The intensity of line emissions depends on the element constituent and the temperature in the atmosphere. The dominance of helium lines over hydrogen lines and their weakening at later times can be due to the elemental constituent density or the optical depth. \citet{2016ApJ...827....3R} suggested that a main sequence star disruption can also result in TDEs with helium dominance due to the optical depth.

Given the current state of modelling, we aim to explore the evolution of the atmosphere to understand the time evolution of hydrogen and helium line luminosity. We look for the elemental abundances along with the optical depth necessary to explain the observations and the nature of the disrupted star. For this study, we use a TDE source iPTF16axa because of the availability of multi-wavelength data on several epochs and the significant helium and hydrogen line luminosity, and their evolution with time. In the spectrum of the TDE iPTF16axa, helium lines dominate over hydrogen lines and the optical/UV luminosity dominates over the X-ray luminosity.  

In this paper, we perform independently the disc-wind model fit to the iPTF16axa spectra continuum and the CLOUDY modelling to the hydrogen and helium line luminosities. Using CLOUDY modelling, we estimate the atmosphere properties which we assume are due to the outflowing wind from a super-Eddington disc and the atmosphere element constituent which helps to explore the possibility of stellar type. We use the disc-wind model to fit the spectrum continuum obtained using optical/UV photometry to interpret the atmosphere from the perspective of TDE disc accretion inflow and outflow. This independent analysis helps in understanding the outflow dynamics that constitute the atmosphere. The independent analysis constrains the outflow properties such as density, temperature, and velocity, and is useful to probe the type of the disrupted star. We show that the CLOUDY modelling of the line luminosity and the disc-wind modelling of the continuum shows comparable results. The continuum modelling shows that the TDE is from the disruption of a high mass star, and the modelling of spectral line luminosities requires a super-solar abundance of helium. We use the disc-wind model as the term to represent disc-wind modelling to continuum spectrum and CLOUDY modelling as the term to represent CLOUDY modelling to spectral lines. 

In section \ref{obs}, we present the details of the iPTF16axa observations. In section \ref{dwmodel}, we present the disc-wind model used to fit the observations and the results of the model fit to the spectrum continuum. In section \ref{cloudy}, we present the analysis of photoionization modelling of CLOUDY to match the observed line luminosity. We compare the results of CLOUDY modelling with the disc-wind modelling in section \ref{results}. The discussion of various aspects of the modelling is in section \ref{discussion} and a summary is presented in section \ref{summary}.

\section{Observational details}
\label{obs}

The source iPTF16axa was discovered on 2016 May 29, by the intermediate Palomar Transient Factory (iPTF) at a redshift of $z=0.108$ \citep{2017ApJ...842...29H}. We select this TDE because of the multi-wavelength data in both spectral continuum and lines
available for several epochs so that the joint analysis using both the continuum and lines can be done. This particular TDE has line luminosity data for several epochs for the CLOUDY modelling. The multi-wavelength observations were done in six filters of SWIFT UVOT that include UVW2(1928 \AA), UVM2 (2246 \AA), UVW1 (2600 \AA), U (3465 \AA), B (4392 \AA) and V (5468 \AA). The blackbody fit to the transient spectral energy distributions (SEDs) obtained using the optical/UV photometry resulted in a temperature of $\sim 3 \times 10^4 ~{\rm K}$. 

The $0.3-10~{\rm keV}$ X-ray luminosity of iPTF16axa using a power law spectrum of photon index $\Gamma = 2$ is $L_X < 3.3 \times 10^{41}~{\rm erg~s^{-1}}$. The single blackbody spectrum temperature from X-ray observation is $T \sim 10^5 ~{\rm K}$. The luminosity in the UV band is higher than the X-ray luminosity and the two temperatures of $\sim 10^5~{\rm K}$ and $\sim 10^4~{\rm K}$ agree with the disc-wind model and reprocessing of X-ray emission to shorter wavelengths. We use the SEDs from optical/UV photometry obtained on days 14, 19, 34, 39, 48, 50, 55, 57, 62, 71, 76, and 83 after detection. We simultaneously fit the disc-wind model to the observed continuum on all days. We thank Tiara Hung for providing photometric data corresponding to Fig 5. in \citet{2017ApJ...842...29H} for continuum modelling. The line luminosity for hydrogen and helium lines are given in \citet{2017ApJ...842...29H}. The spectroscopy data is available at \url{https://wiserep.weizmann.ac.il/object/575}.

The spectrum is dominated by the Helium line HeII at the initial time and the H$\alpha$ line was readily detected except for the last epoch. The iPTF16axa did not show significant H$\alpha$ suppression as was observed in PS1-10jh. The luminosity ratio of HeII to H$\alpha$ lines decreases from $\sim 2$ to $\sim 0.4$ over a period of $~ 80$ days, but the blackbody temperature shows weak variations with a mean value at $\bar{T} = 3.0 \pm 0.33 \times 10^4 ~{\rm K}$. The FWHM inferred from He II and H$\alpha$ lines is $\sim 10^4~{\rm km~s^{-1}}$. The linewidth of He II remains comparable, and sometimes even narrower than the H$\alpha$ throughout the observations. 

In the next section, we present the CLOUDY analysis of the spectrum of iPTF16axa.

\section{disc-wind modelling}
\label{dwmodel}

In this section, we present the basics of disc-wind modelling and the fit to the transient SEDs. The dynamics of an outflow from the accretion disc and their emission is still an unsolved problem in TDEs. However, there are various emission models proposed to explain the observations. \citet{2019MNRAS.483..565C} through GRMHD simulations have shown that the disc in a radiation pressure-dominated regime with weak magnetic field results in a radiation-driven outflowing wind. In the introduction, we have presented the various accretion and outflow emission models. The outflow geometry is complex but the basic formalism for a spherical outflow has been laid by the earlier works \citep{2009MNRAS.400.2070S,2020ApJ...894....2P,2020ApJ...905L...5U}. We employ the spherical wind-driven outflow model of \citet{2020ApJ...905L...5U} with the fraction of mass outflow, $f_0$, varying with time instead of being constant. We use the disc-wind model to fit the continuum and deduce the effective temperature in the disc and wind necessary to formulate the initial spectrum in the CLOUDY. A more detailed wind-driven model for the TDEs spectrum and the emissions will be taken up in the future but here, we focus on using the CLOUDY to estimate the atmosphere properties by fitting the observed line luminosities and the comparison of the CLOUDY results to the estimation from the disc accretion inflow and outflow model.   

\subsection{Models}
\label{dwmodel_theory}

The tidally disrupted debris returns to the pericenter following an orbit where it interacts resulting in shocks that circularizes the infalling debris and form an accretion disc. The formation of an accretion disc depends on the momentum exchange of debris streams and the radiative efficiency of the heat generated due to the viscous mechanism and internal energy of the debris. If the circularization is radiatively efficient, the formed disc is a thin disc, or else a thick disc is formed \citep{2016MNRAS.461.3760H}. A thick disc is dominated by advection with radiation pressure and an outflow is launched when the radiative flux exceeds the critical value governed by gravity. In a hydrostatic equilibrium with vertical radiation flow, the outflow exists for radiative flux $Q_{r}^{-} \gtrsim \mathcal{F}_{\rm max} = (2 \sqrt{3}/9)(c/\kappa) G M_{\bullet}/r^2$, where $c$ is the speed of light, $\kappa$ is the opacity in the medium, $G$ is the gravitational constant, $M_{\bullet}$ is the black hole mass and $r$ is the radial distance \citep{2015MNRAS.448.3514C}. These authors showed that the critical mass accretion rate above which the outflow exists is $\dot{M}_{\rm crit} \sim 1.04 \dot{M}_E$, where the Eddington rate is calculated for the efficiency of 0.1. If we neglect the advection and assume the energy generated via viscosity equates to the radiative loss, then following the  standard Shakura-Sunyaev accretion model, the radiative flux is $Q_r^{-} = (3/4 \pi) G M_{\bullet} \dot{M}_a f/r^3$, where $f = 1 - \sqrt{r_{\rm in}/r}$ with the disc inner radius $r_{\rm in}$. Assuming $r \gg r_{\rm in}$, the criteria for outflows results in $r < r_{\rm crit} = (9 \sqrt{3}/8 \pi) (\kappa/c) \dot{M}_a$. This implies that the radiative flux is inefficient to induce an outflow for a radius greater than $r_{\rm crit}$. From the angular momentum conservation, the circularization radius of TDE discs is $r_c = 2 r_t$, and thus the entire disc has an outflow if $ r_c \leq r_{\rm crit}$, which results in $\dot{M}_a \geq 0.602 (\eta / 0.1) (M_{\star}/M_{\odot})^{7/15} [M_{\bullet}/(10^6 M_{\odot})]^{-2/3} \dot{M}_E$. The presence of advection energy loss reduces the radiative energy loss and can suppress the outflows. Thus, the critical mass accretion rate increases compared to the case without advection. 

The presence of advection results in a slim disc where the disc height $H \sim r$ and the pressure is dominated by radiation pressure. In a steady state disc, the mass conservation results in the total mass loss rate in the disc which is the sum of the mass accretion rate ($\dot{M}_a$) and the mass outflow rate ($\dot{M}_{\rm out}$) to be a constant and is taken to be the mass fallback rate $\dot{M}_{\rm fb}$. SQ2009 constructed a steady structure slim disc with constant mass accretion rate given by $\dot{M}_a = (1 - f_0) \dot{M}_{\rm fb}$, where $f_0$ is a constant. The radiative flux obtain is given by 

{\footnotesize
\begin{equation}
Q_r^{-} = \sigma_{\rm SB} T_{\rm eff}^4 = \frac{3 G M_{\bullet} \dot{M}_a f}{8 \pi r^3} \left[\frac{1}{2}+ \sqrt{\frac{1}{4}+ \frac{3}{2} f \left(\frac{\dot{M}_a}{\eta \dot{M}_E}\right)^{2} \left(\frac{r}{r_s}\right)^{-2}}\right]^{-1},
\label{tefdisc}
\end{equation}
}

\noindent where $r_s = 2 G M_{\bullet}/c^2$ is the Schwarzschild radius. Using this in $Q_{r}^{-} \gtrsim \mathcal{F}_{\rm max}$, we obtain $r < (49/81) (9 \sqrt{3}/8 \pi) (\kappa/c) \dot{M}_a$, where we assume $r \gg r_{\rm in}$. Thus, we can see that the critical radius $r_{\rm crit}$ decreases by 0.605. Here, the requirement for the entire disc to be in the outflow phase results in $\dot{M}_a \geq 0.995 (\eta / 0.1) (M_{\star}/M_{\odot})^{7/15} [M_{\bullet}/(10^6 M_{\odot})]^{-2/3} \dot{M}_E$. The parameter $f_0$ is not a constant and its evolution depends on the disc dynamics. \citet{2011MNRAS.413.1623D} numerically simulated the fraction of mass outflow in terms of mass accretion rate at the outer radius for a steady disc and the numerical result is approximated into empirical relation given by $f_0 = (2/\pi) {\rm Tan^{-1}}[(\dot{M}_{\rm fb}- \dot{M}_E)/(4.5 \dot{M}_E)]$ (MM15). The mass fallback rate of the infalling debris is taken to be

\begin{equation}
\dot{M}_{\rm fb} = \frac{1}{3}\frac{M_{\star}}{t_m} \left(\frac{t_m + t}{t_m}\right)^{-5/3},
\label{mfb}
\end{equation}

\noindent where $M_{\star}$ is the stellar mass and the orbital period of innermost bound debris is given by

\begin{equation}
t_m = 40.8~{\rm days}~ \left(\frac{M_{\bullet}}{10^6 M_{\odot}} \right)^{1/2} \left(\frac{M_{\star}}{M_{\odot}}\right)^{1/5} k^{-3/2},
\end{equation}

\noindent where the radius of star $R_{\star}=R_{\odot} (M_{\star}/M_{\odot})^{0.8}$ \citep{1994sse..book.....K}, and $k$ is the tidal spin-up factor to take into account the spin-up of a star due to the tidal torque by the black hole \citep{2001ApJ...549..948A,2020MNRAS.496.1784M}. The numerical simulations by \citet{2001ApJ...549..948A} showed that the tidal interactions lead to an energy transfer from the orbit to the star that spin-up the star and by including the tidal interactions, \citet{2002ApJ...576..753L} formulated the disruption dynamics and obtained $k = 3$. There would be some deviation at the initial time due to stellar structure \citep{2009MNRAS.392..332L}, but we use the equation (\ref{mfb}) as the initial rise is for only a few days and the power-law decline dominates most of the evolution phase. 

\begin{figure}
\centering
\includegraphics[scale = 0.48]{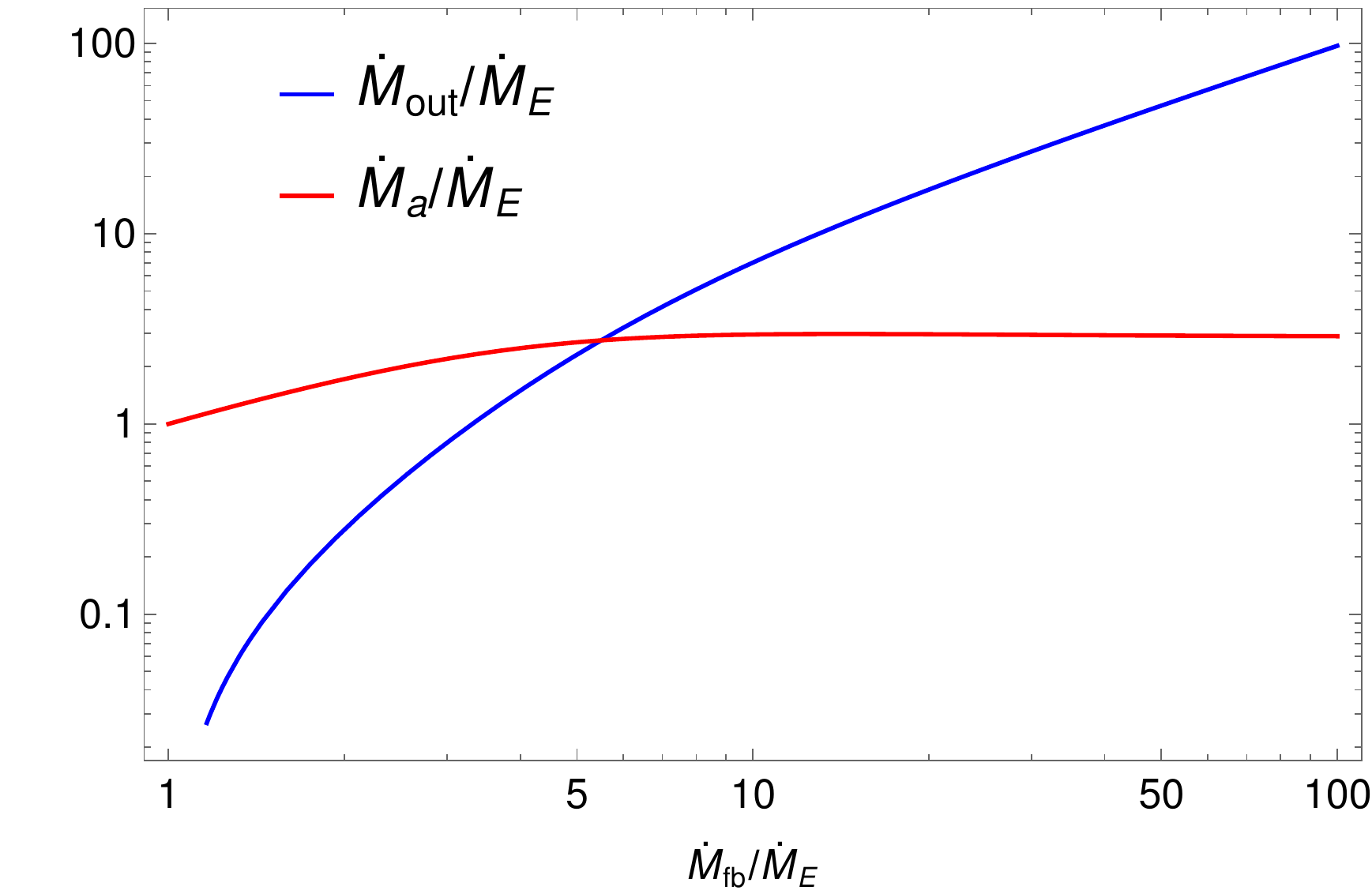}
\caption{The mass accretion rate ($\dot{M}_a$: red line) and the mass outflow rate ($\dot{M}_{\rm out}$: blue line) are shown as a function of the mass fallback rate (which is the total mass loss rate from the steady disc, $\dot{M}_a$ + $\dot{M}_{\rm out}$). See section \ref{dwmodel_theory}. }
\label{mfnt}
\end{figure}

The dynamical geometry and the evolution of wind in TDEs are not yet clear. However, we model a steady-state wind structure to fit the SEDs continuum obtained using optical/UV bands (see Fig. 5 of \citealp{2017ApJ...842...29H}) and explore the evolution of atmosphere properties due to the wind. We assume a spherical outflowing wind with wind persisting at all radii at a time. The wind evolves with a constant velocity $v_w$ and the mass outflowing rate $\dot{M}_{\rm out}$. The density profile in the wind is given by 

\begin{equation}
\rho(r) = \frac{\dot{M}_{\rm out}}{4 \pi r^2 v_w}, 
\label{dens}
\end{equation}

\noindent where $\dot{M}_{\rm out} = f_0 \dot{M}_{\rm fb}$. Fig. \ref{mfnt} shows the evolution of mass outflow rate and mass accretion rate as a function of the mass fallback rate. The mass outflow rate dominates when the mass fallback rate is high. The wind is launched at a radius $r_{l}$ which is also assumed to be the inner radius of the wind. We assume that the outer radius of the wind $r_{\rm w} \gg r_{l}$. The optical depth $\tau$ for an opacity dominated by electron scattering is given by 

\begin{equation}
\tau(r) = \int_{r}^{r_w} \kappa_{\rm es} \rho(r) \, \diff r = \frac{\kappa_{\rm es} \dot{M}_{\rm out}}{4 \pi v_w} \left(\frac{1}{r} -\frac{1}{r_w}\right).  
\label{taur}
\end{equation}

\noindent In the limit $r_{\rm w} \gg r_{l}$, the optical depth is given by $\tau(r) = \kappa_{\rm es} \dot{M}_{\rm out}/(4 \pi v_w r)$. Following \citet{2009MNRAS.400.2070S}, we consider that at the inner radius $r_l$, the kinetic energy of the outflow [$(1/2)\rho(r) v_w^2$] is in equilibrium with the internal energy dominated by the radiation ($a T_l^4$, where $a$ is the radiation constant and $T_l$ is the temperature at $r_l$). Thus, the temperature at the inner wind is given by

\begin{equation}
T_l  = \left(\frac{\dot{M}_{\rm out}v_w}{4 \pi r_{l}^2 a} \right)^{1/4}.   
\end{equation}

The outflowing wind near the inner wind radius is dominated by radiation pressure and if the time of photon diffusion is higher than the dynamical time of the outflowing wind, the photons are trapped and coupled with the matter. The photon diffusion timescale is given by \citep{2020ApJ...894....2P}

\begin{equation}
t_{\rm diff} = \frac{\tau(r)}{c} \frac{r(r_w - r)}{r_w},
\end{equation}

\noindent at the dynamical time is given by $t_{\rm dyn} = (r - r_l)/v_w$. Taking $r_{\rm w} \gg r_{l}$ and using equation (\ref{taur}), the radius up to which the photon is trapped also known as trapping radius ($r_{\rm tr}$) can be obtained using $t_{\rm diff} = t_{\rm dyn}$, and is given by 

\begin{equation}
r_{\rm tr} = r_l + \frac{\kappa_{\rm es} \dot{M}_{\rm out}}{4 \pi c}.
\label{rtr}
\end{equation}

\noindent Within the trapping radius, $r < r_{\rm tr}$, the photons are advected and the temperature evolves adiabatically. The adiabatic temperature is given by $T \propto \rho(r)^{1/3} \propto r^{-2/3}$. Thus, the temperature at the trapping radius is given by 

\begin{equation}
T_{\rm tr} = T_l \left(\frac{r_{\rm tr}}{r_l}\right)^{-2/3}.
\end{equation}

Above the trapping radius, $r > r_{\rm tr}$, the photons diffuse through the wind such that the luminosity is given by 

\begin{equation}
L = -\frac{4 \pi r^2 a c}{3 \kappa_{\rm es} \rho(r)} \frac{\partial T^4}{\partial r}.
\label{lumd}
\end{equation}

\noindent The luminosity is nearly constant in the diffusive regime and thus, using equation (\ref{dens}), the temperature evolves as $T \propto r^{-3/4}$. Using equation (\ref{lumd}), the luminosity at the trapping radius is given by 

\begin{equation}
L = \frac{2 \pi c}{3 \kappa_{\rm es}} v_w^2 r_l \left(\frac{r_{\rm tr}}{r_l}\right)^{1/3}.
\label{lwind}
\end{equation}

In the diffusive region, apart from electron scattering, the absorption of photons by the electrons can also be a source of opacity. Thus, total opacity is given by $\kappa_{\rm eff} = \sqrt{3 \kappa_{\rm es} \kappa_a}$, where the absorption opacity is taken to be the Kramers opacity given by $\kappa_a = \kappa_0 \rho T^{-7/2} ~{\rm cm^{2}~g^{-1}}$, assuming density and temperature in cgs units \citep{2020ApJ...905L...5U}. The colour radius, $r_{\rm cl}$, where the optical depth  $\int_{r_{\rm cl}}^{r_w} \kappa_{\rm eff} \rho(r) \, \diff r =1$, is given by 

\begin{equation}
r_{\rm cl} = \left(\frac{16}{11}\right)^{16/11} \left(3 \kappa_{\rm es}\kappa_0\right)^{8/11} \left(\frac{\dot{M}_{\rm out}}{4 \pi v_w}\right)^{24/11} r_{\rm tr}^{-21/11} T_{\rm tr}^{-28/11}.
\end{equation}

The photospheric radius is given by 

\begin{equation}
r_{\rm ph} = {\rm Max [} r_{\rm tr},~r_{\rm cl} {\rm]},
\end{equation}

\noindent and the corresponding effective temperature of emission is given by 

\begin{equation}
T_{\rm ph} = \left(\frac{L}{4 \pi r_{\rm ph}^2 \sigma}\right)^{1/4}.
\label{tpht}
\end{equation}

\noindent where $\sigma$ is the Stefan-Boltzmann constant. Assuming a blackbody emission, the luminosity in a given spectral band $\{\nu_l,~\nu_h\}$, is given by

\begin{equation}
L_{\rm \{\nu_l,~\nu_h\}} = 4 \pi r_{\rm ph}^2 \int_{\nu_l}^{\nu_h} \frac{2 h \nu^3}{c^2} \frac{1}{Exp[\frac{h \nu}{k_B T_{\rm ph}}]-1} \, \diff \nu.
\label{lwnu}
\end{equation}

\noindent The spectral luminosity at a given wavelength is given by

\begin{equation}
\lambda L_{\lambda} = \lambda ~4 \pi r_{\rm ph}^2  \frac{2 h c^2}{\lambda^5} \frac{1}{\exp[\frac{h c}{\lambda k_B T_{\rm ph}}]-1}
\label{lspec}
\end{equation}

\subsection{Fit to data}
\label{fitdw}

We consider the spectra of the source iPTF16axa (section \ref{obs}) on various days. The spectra using optical/UV magnitudes are given as Fig. 5 in \citet{2017ApJ...842...29H} and were fit by a single blackbody temperature model.   We assume that the optical/UV emissions are from the wind and fit equation (\ref{lspec}) to the spectrum on individual days. We take the black hole mass to be $M_{\bullet} = 2.18 \times 10^6 M_{\odot}$ \citep{2017MNRAS.471.1694W}. We introduce a time parameter $\Delta t$ such that time $t$ in equation (\ref{mfb}) is $t + \Delta t$. The tidally disrupted debris circularizes to form an accretion disc and the circularization time for the debris is unknown. We assume that the total mass loss rate from the disc (sum of mass loss due to the black hole accretion and outflow) follows the mass fallback rate of the debris. The $\Delta t$ is the shift in time $t$ to consider the uncertainty in the starting time of disc accretion after the innermost debris returns to the pericenter and is useful in fitting the continuum. By fitting, we estimate the values of parameters such as stellar mass $M_{\star}$, radiative efficiency $\eta$, wind inner radius $r_l$, constant wind velocity $v_w$ and $\Delta t$. The other parameters such as density, temperatures, and radius of the photosphere depend on these free parameters and are obtained later after the fit.
 
We perform a $\chi^2$ minimization simultaneously on all observational data points. Using the obtained parameters, we calculate the Fisher-Information matrix by taking the second derivative of log-likelihood with respect to the free parameters. The inverse of the Fisher matrix is the covariance matrix and the square root of diagonal elements is the standard error of parameters.

The parameters obtained are $M_{\star} = 6.20 \pm 1.19 M_{\odot}$, $\log_{\rm 10}(\eta) = -1.222 \pm 1.327$, $r_l = (2.013 \pm 0.551) \times 10^{14}~{\rm cm}$, $v_w = 18999.4 \pm 1785.1 ~{\rm km~s^{-1}}$, and $\log_{\rm 10}(\Delta t~{\rm (days)}) = 0.491 \pm 0.5$, with a reduced $\chi^2 = 2.44$. The significant standard errors in $\eta$ and $\Delta t$ imply that these values are weakly constrained.

\begin{figure*}
\centering
\includegraphics[scale = 0.5]{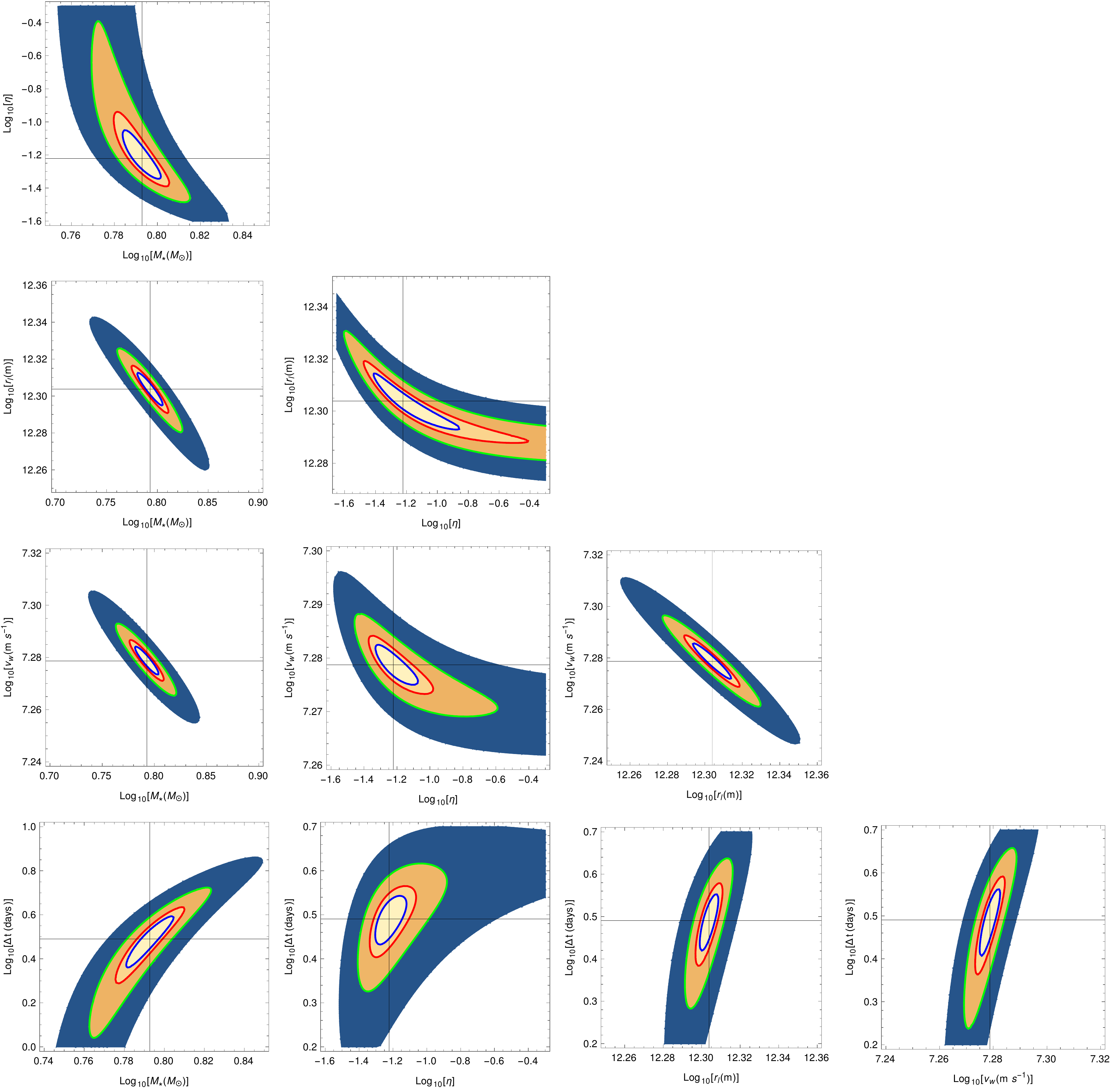}
\caption{The density plots show the relative likelihood $\bar{\mathcal{L}} = \mathcal{L} / \mathcal{L}_p > 0.1$, where $\mathcal{L}_p$ is the likelihood at the obtained parameters values which are shown by solid black lines. The blue, red, and green lines represent the relative likelihood $\bar{\mathcal{L}} =$ 90\%, 80 \% and 50 \%, respectively. The obtained parameters are within 90\% of the maximum likelihood. See section \ref{fitdw}}
\label{cnt}
\end{figure*}

\begin{figure*}
\centering
\includegraphics[scale = 0.57]{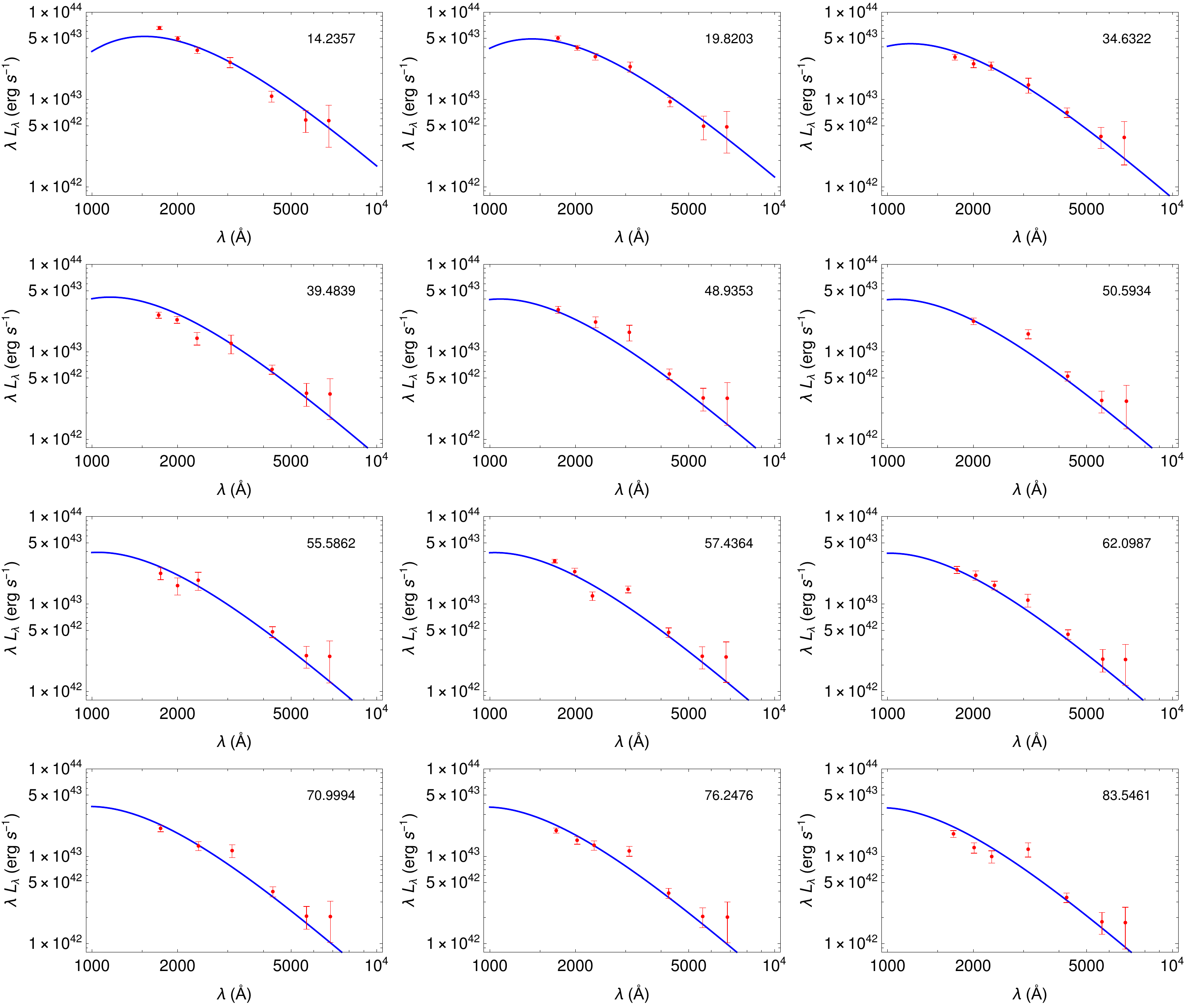}
\caption{The continuum spectrum from optical/UV observations of iPTF16axa (red) are shown along with the model spectra (blue) using the disc-wind model presented in section \ref{dwmodel_theory} and the luminosity given by equation (\ref{lspec}) for the obtained parameters. The corresponding day of observation is shown at the top right of each panel. The obtained parameters are the stellar mass $M_{\star} = 6.20 \pm 1.19 M_{\odot}$, disc radiative efficiency $\log(\eta) = -1.222 \pm 1.327$, wind inner radius $r_l = (2.013 \pm 0.551) \times 10^{14}~{\rm cm}$, wind velocity $v_w = 18999.4 \pm 1785.1 ~{\rm km~s^{-1}}$, and $\log(\Delta t~{\rm (days)}) = 0.491 \pm 0.5$. See section \ref{fitdw}.}
\label{lspecplt}
\end{figure*}

Using the obtained parameter values, we generate a relative likelihood $\bar{\mathcal{L}} = \mathcal{L} / \mathcal{L}_p$, where $\mathcal{L}$ is the likelihood and $\mathcal{L}_p$ is the likelihood at the obtained parameters. Fig. \ref{cnt} shows the relative likelihood contours around the obtained parameters and the obtained parameters are within 90\% of the peak likelihood $\mathcal{L}_p$. The blue region represents the relative likelihood $\bar{\mathcal{L}} > 0.1$ and the region is distributed over a large range for $\eta$ and $\Delta t$. Thus, the standard errors are substantial and imply a weak constraint on these parameters. Fig. \ref{lspecplt} shows the model spectra at the obtained parameters with the observed optical/UV continuum on various days.

\section{Modelling using CLOUDY}
\label{cloudy}

We use the photoionization code CLOUDY (c17.02) \citep{2017RMxAA..53..385F}\footnote{\url{https://www.nublado.org}} to model the optical spectral lines, He II 4685.64 \AA ~ and H$\alpha$ (H I 6562.81 \AA), from iPTF16axa at four different epochs \citep{2017ApJ...842...29H}, and derive the underlying physical conditions that gave rise to them. The numerical simulation code CLOUDY is based on microphysics. It simulates the thermal, ionization, and chemical
structure of a cloud over a wide range of physical conditions and predicts its observed spectrum 
with the help of a very few input parameters like density, chemical composition, radiation field etc., and vice versa. 
It is used in various astrophysical environments where non-equilibrium astrophysical plasma is involved \citep{1992ApJ...387...95F, 1995ApJ...455L.119B, 2006ApJ...639..941S, 2008ApJ...675..405S, 2009ApJ...701..677S, 2016MNRAS.459.3234S, 2018MNRAS.481.2083R, 2018MNRAS.474.3307S, 2019MNRAS.486..195S, 2019MNRAS.483.4884M, 2020MNRAS.493.5153S, 2021MNRAS.507.1441S, 2021ApJ...920L..25S, 2022ApJ...934...53S, 2022ApJ...925..187P}. 
Details about CLOUDY can be found in \citet{2013RMxAA..49..137F, 2017RMxAA..53..385F} and references given there.

This is our first attempt to model TDE emission lines using CLOUDY. However, many researchers have used CLOUDY to model TDE emission lines. \citet{2018MNRAS.474.3307S} showed that standard AGN-like irradiation models for TDEs cannot explain both the line ratios and the individual 
equivalent widths seen in
observations simultaneously. Recently \citet{2021ApJ...920L..25S} have modelled the CIV/NIV] and ratios between different nitrogen lines from GSN 069 using CLOUDY. They argued that the abnormal C/N abundance ratio is due to a partially disrupted red giant 
star. However, the line profiles were not well reproduced.
Following these, we use blackbody irradiations (with luminosity predicted by our disc-wind model
calculations as discussed in the previous section) to model both the He/H abundance ratio and the line profiles from iPTF16axa 
to decipher the disruptive star’s type. 

Here, we perform a grid of models in the parameter space (See section \ref{cloudy_parameters}). We then choose the parameter values that predict a better match to the observed values and finally fine-tune the relevant input parameters so that the observed data can 
be optimally modelled.

\subsection{Models and input parameters}
\label{cloudy_parameters}
We assume a spherical atmosphere of ionized gas expanding with a constant velocity of $\textit v~{\rm km~s^{-1}}$ and 
the dimension of this gas is defined by an inner radius $r_{l}~{\rm (cm)}$ and a thickness dependent on the time elapsed multiplied by $\textit v$. Earlier, a similar spherical model was used by \citet{2016ApJ...827....3R}. The
observed FWHM of the He II and H$\alpha$ lines of iPTF16axa reveal an expanding velocity $\approx$ 10$^4$ km s$^{-1}$ for the four epochs \citep{2017ApJ...842...29H}. We vary the free parameter $\textit v$ and our final value, 
$1.3 \times 10^{4}~{\rm km~s^{-1}}$, is close to the observed value \citep{2017ApJ...842...29H}. In all our models presented here, we assume a density law proportional 
to $\rho_{l} r^{-2}$ \citep{2016ApJ...827....3R}, where $\rho_{l}~{\rm (cm^{-3} )}$ is the total hydrogen number density at the inner radius $r_{l}~{\rm (cm)}$. Both $r_{l}~{\rm (cm)}$ and $\rho_{l}~{\rm (cm^{-3} )}$ are free parameters. Though $r_{l}$ remains the same for all four epochs, the values of $\rho_{l}$ are different. Following the optical observations and \citet{2020SSRv..216..114R}, we assume the gas to be irradiated at $r_{l}~{\rm (cm)}$ by two blackbody radiations with temperatures T$_{BB1}$ and T$_{BB2}$ K, arising from the disc and the wind, respectively. T$_{BB1}$ and T$_{BB2}$ K are free parameters. However, like $\textit v$, we vary T$_{BB1}$ and T$_{BB2}$ close to the observed values, $10^{5}~{\rm K}$ and $10^{4}~{\rm K}$, respectively. Furthermore, we try to keep the wind and disc luminosities within the range predicted by our disc-wind model calculations described in section \ref{dwmodel}. We use solar abundance \citep{2010Ap&SS.328..179G} for all the elements except He. The He abundance is a free parameter. 
 
Our models are moderately optically thick to electron scattering. So, while varying the parameters we had to consider only cases within the permissible range of electron scattering optical depths internally set by CLOUDY.

CLOUDY can calculate both time-dependent and time-independent models. However, the time-dependent calculations are a thousand times slower than time-steady calculations. Hence, in this work, we do not use any explicit time dependence. Instead, we use four different time-independent non-local Thermodynamic Equilibrium (NLTE) snapshot models for the four epochs and reproduce the observed luminosities of HeII and HI lines at those epochs. This saves significant computational time compared to a time-dependent model. This is justified as various formation and recombination timescales are smaller compared to the age of the wind. Similar static snapshot models at different epochs were done earlier to study the time evolution of various astrophysical environments \citep{1989A&A...213..274S,2020A&A...639L..12S,2022ApJ...925..187P}.

\subsection{CLOUDY Results}
\label{cloudy_res}
Table \ref{tab:table 1} shows the physical parameters for our final model of iPTF16axa using CLOUDY. 
Our final model estimates $r_{l} = 10^{14.85}~{\rm cm}$. We notice 
that the value of $\rho_{l}~{\rm (cm^{-3} )}$ and wind luminosity slightly decrease as time passes. 
In addition, our final models predict a much smaller optical depth for the He II line compared to the H$\alpha$ line.
Despite this, our final models require super solar abundance for He. Hence, we conclude that both the super solar abundance of He 
and a smaller He II line optical depth are responsible for the enhancement of He II line luminosity over the H$\alpha$ line luminosity.
The surface abundance of He depends both on the stellar mass and age.    
In our case, a three times solar abundance of He reproduces observed line luminosities for all the epochs. It suggests that the star that got disrupted most probably is an evolved red giant \citep{2019ApJ...878...49W}. The abundance of other elements could also be derived if lines from those elements were observed. 

\begin{table}
	\centering
	\caption{Input  parameters for our best models for four epochs of iPTF16axa.}
	\label{tab:table 1}
	\begin{tabular}{llllll} 
		\hline
	Days& $r_{l}$&$\rho_{l}$  & L(T$_{BB1}$=10$^{5}$ K) & L(T$_{BB2}$=10$^4$ K)\\          
	&log (cm)&log (cm$^{-3}$) & log (erg s$^{-1}$) & log (erg s$^{-1}$) \\ 
		\hline
		5&14.85&  10.01 &42.86 &44.14\\
	    11&14.85& 10.00  &42.90 &44.12\\
	    14&14.85& 9.95  &42.90 &44.10\\
	    37&14.85& 9.91 &42.85 &43.97\\
		\hline
		\end{tabular}
        \end{table} 

\begin{figure}
    \centering
    \includegraphics[scale=0.6]{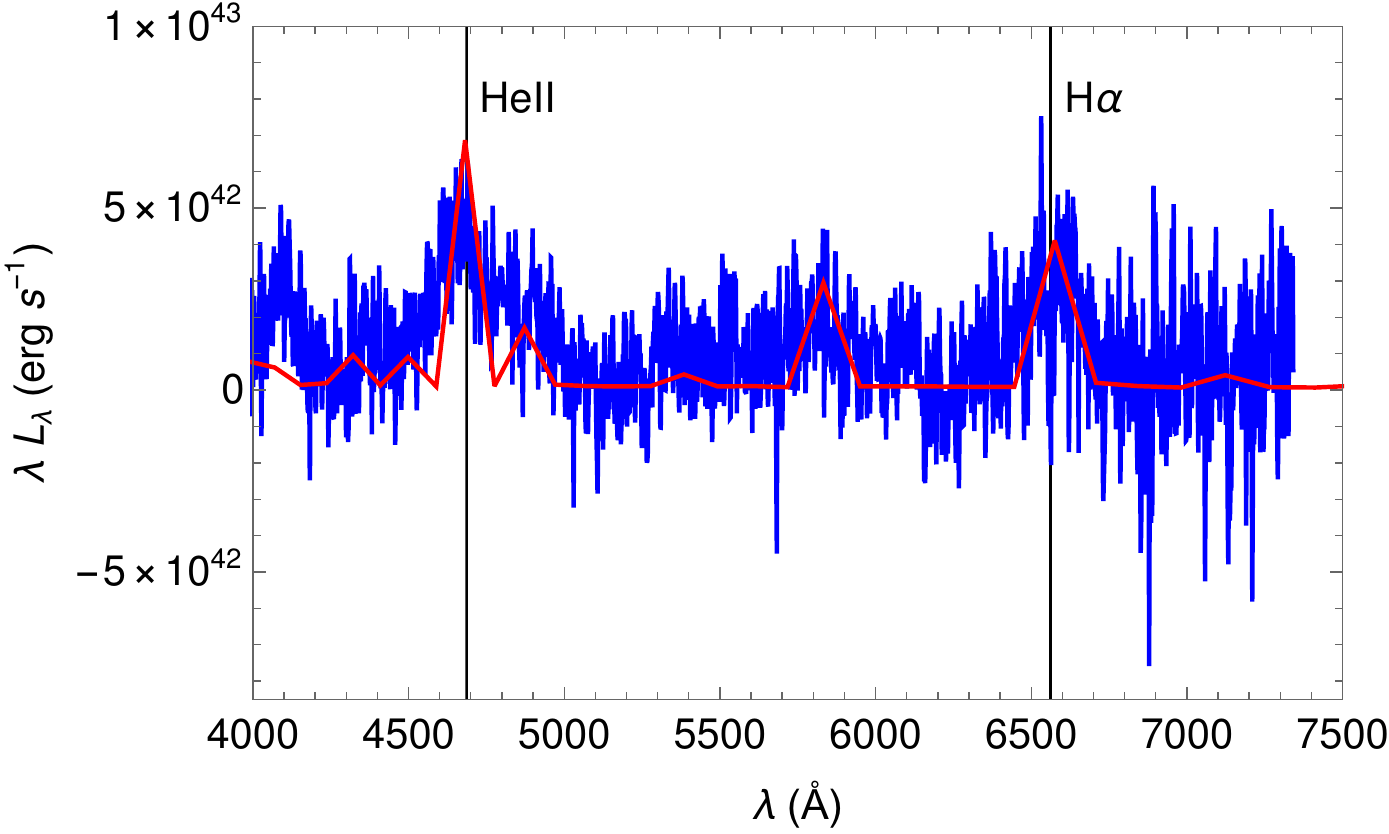}
    \caption{The red line, which is the continuum subtracted CLOUDY spectrum, shows helium and hydrogen lines for day 14. This is compared with the continuum subtracted observed spectrum shown in blue. See section \ref{cloudy} for details.}
    \label{fig:spectrum_fit}
\end{figure}

\begin {figure}
\includegraphics[scale=0.5]{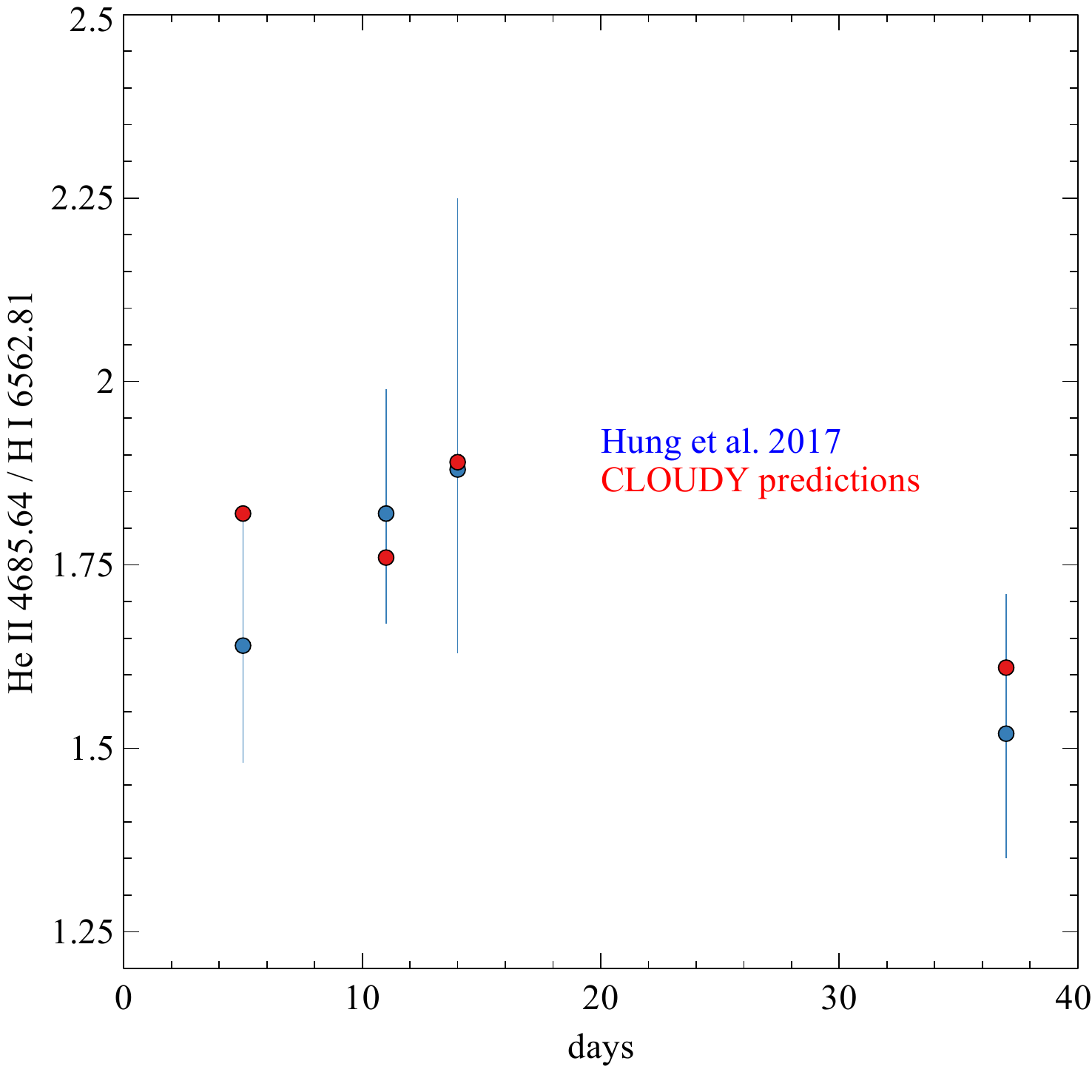}
\caption{The ratio of helium to hydrogen line luminosity from the observations (blue) and the CLOUDY modelling (red) is shown at various epochs.}
\label{fig:tde_ratio_2022}
\end {figure}

Table \ref{tab:table 2} compares the observed and the model-predicted line luminosities of our final models for four epochs. 
Our model-predicted line luminosities 
of He II 4685.64 \AA ~ and H$\alpha$ match with observation. Fig. \ref{fig:spectrum_fit} shows the continuum subtracted spectrum from CLOUDY and the observation for day 14. Fig.~\ref{fig:tde_ratio_2022} compares the observed and model-predicted line ratios using CLOUDY. The vertical blue lines represent observed error bars.

\begin{table}
	\centering
	\caption{Comparison of the observed and the model predicted line luminosities for 4 epochs using CLOUDY (in units of 10$^{40}$ erg s$^{-1}$)}
	\label{tab:table 2}
	\begin{tabular}{ccccc} 
		\hline
	Days& He II 4685.64& H$\alpha$  &He II 4685.64 \AA &H$\alpha$ \\          
	&This work & This work &Observed& Observed\\ 
		\hline
		5&  13.3 & 7.3 & 12.5$\pm$0.8& 7.6$\pm$0.3\\
	    11& 14.6 & 8.3 & 15.1$\pm$0.6 & 8.3$\pm$0.4\\
	    14& 12.3 & 6.5 &12.8$\pm$0.7  & 6.8$\pm$0.8\\
	    37& 9.7 & 6.01 & 9.1$\pm$0.6 & 6.0$\pm$0.3\\
		\hline
		\end{tabular}
        \end{table}
        
CLOUDY calculates gas temperature self-consistently from heating and cooling balance consisting 
of various physical processes \citep{2013RMxAA..49..137F}. The total heating (and total cooling) for the four epochs are 10$^{42.329}$, 10$^{42.413}$, 10$^{42.334}$, and 10$^{42.264}$ ergs s$^{-1}$, respectively. 

The dominant heating processes are the ionization of He and H. Whereas, the cooling is dominated by O VI, O V, CIV, 
C III, and N V lines. The electron temperature averaged over the thickness of the ionized gas for the four epochs 
are 1.77$\times$10$^4$, 1.73$\times$10$^4$, 1.71$\times$10$^4$, 1.65$\times$10$^4$ K, respectively. The gas is fully ionized. Hydrogen and Helium are mainly in H$^+$, He$^{+}$, He$^{++}$. The physical conditions described above are suitable to 
form molecular ion HeH$^+$. Our CLOUDY simulations do produce HeH$^+$ but the predicted column density is below 10$^{10}$ cm$^{-2}$. In Table \ref{tab:table 3}, we list the predicted column densities of some important 
ions with column density greater than 10$^{16.5}$ cm$^{-2}$ for all four epochs. 

\begin{table}
	\centering
	\caption{The estimated Column densities in log (cm$^{-2}$) for iPTF16axa using CLOUDY}
	\label{tab:table 3}
	\begin{tabular}{ccccc} 
		\hline
	Species & Day 5& Day 11 & Day 14 & Day 37\\          
		\hline
		H I & 17.665  & 17.757&17.627 & 17.674\\
	    H II & 24.609 & 24.713& 24.686& 24.697\\
	    He I& 16.576 & 16.649& 16,521& 16.607\\
	    He II& 23.855 & 23.991& 23.936& 24.005\\
	    He III & 23.755&23.817 &23.828 & 23.749\\
	    C III &17.140 &17.372 & 17.238& 17.334\\
	    C IV &21.006 &21.114 & 21.081 & 21.096\\
	    C V & 19.901 & 19.953& 20.003& 19.954\\
	    N IV & 20.406 & 20.514& 20.482&20.497\\
	    N V & 18.428 &18.345  & 18.366& 18.347\\
	    O III& 18.985 & 19.284& 19.010&19.274\\
	    O IV & 21.264 & 21.371& 21.340&21.353 \\
	    O V & 20.127 & 20.171& 20.222&20.172 \\
	    Fe IV & 20.174 &20.184&20.151&20.166\\
	    Ne III & 20.603&20.614&20.581&20.596\\
		\hline
		\end{tabular}
        \end{table} 

\subsection{Model limitations}
\label{limit}
These models have some limitations. One important property is that the geometry is unknown
and might be unknowable. In addition to that, our models are restricted within the permissible range of electron 
scattering optical depths internally set by CLOUDY. These can affect the results. Time-dependent CLOUDY modelling 
can be done to gain a better understanding but that is time-consuming.
In future, we aim to investigate  more  such TDEs with enhanced He II/ H$\alpha$ ratios to gain a significant understanding 
of the physical conditions.

\section{Results}
\label{results}

Here, we present the results of the model fit to the continuum spectrum obtained using optical/UV observations and compare them with the CLOUDY results obtained through modelling the spectral lines. Using the disc-wind model to the continuum, the obtained inner radius of the wind is $r_l = (2.013 \pm 0.551) \times 10^{14}~{\rm cm}$, and the wind velocity $v_w = 18999.4 \pm 1785.1 ~{\rm km~s^{-1}}$, which is within the range 5000 to 22387.2 ${\rm km~s^{-1}}$ inferred from the full width half maximum of emission lines (see Fig. 15 in \citealp{2017ApJ...842...29H}). The obtained photosphere radius $r_{\rm ph}$ and the effective temperature $T_{\rm ph}$ are shown in Table \ref{rphtph}. The photosphere radius decreases with time which increases the effective temperature (see equation \ref{tpht}). Our obtained effective temperature $T_{\rm ph}$ is in agreement with the blackbody temperature of $~ (2.6 - 4 ) \times 10^4~{\rm K}$ from the optical/UV luminosity \citep{2017ApJ...842...29H}. 

\begin{table}
\caption{The derived photosphere radius, $r_{\rm ph}$, and the effective temperature, $T_{\rm ph}$, of the wind from the model fit. The single blackbody temperature of the disc, $T_d$, is also shown.}
\label{rphtph}
\scalebox{1.1}{
\centering
\begin{tabular}{c|c|c|c|}
\hline\\
&&&\\
$t~{\rm (days)}$ & $r_{\rm ph}~{\rm ( 10^{14}~ cm)}$ & $T_{\rm ph}~{\rm (10^4~K)}$ & $T_{d} {\rm (10^5~K)}$ \\
&&&\\
\hline \\
&&&\\
5  & $15.22_{-0.86}^{+0.56}$  & $1.99_{-0.34}^{+0.89} $ & $0.95567_{-0.542}^{+1.39}$\\
&&&\\
11 & $11.17_{-6.60}^{+3.89}$ & $2.27_{-0.36}^{+1.02}$ & $0.95603_{-0.543}^{+1.38}$ \\
&&&\\ 
14 & $9.84_{-5.90}^{+3.36}$ & $2.39_{-0.37}^{+1.09}$ & $0.95623_{-0.543}^{+1.37}$\\
&&&\\
37 & $5.22 _{-3.18}^{+1.70}$ & $3.12_{-0.38}^{+1.44}$ & $0.95797_{-0.545}^{+1.25}$ \\
&&&\\
\hline
\end{tabular}
}
\end{table}

The bolometric luminosity of the disc, $L_d$, is calculated assuming a blackbody emission from a slim disc with an effective temperature given by equation (\ref{tefdisc}), where the inner radius is taken to be the innermost stable circular orbit (ISCO with black hole spin $j=0$; \citealp{1972ApJ...178..347B}) and the outer radius is taken to be wind's inner radius $r_l$. The single blackbody temperature of the disc is given by $T_d = [L_d/\{\pi(r_l^2-r_{\rm ISCO}^2) \sigma\}]^{1/4}$. Table \ref{rphtph} shows the $T_d \sim 0.955 \times 10^5 ~{\rm K}$ which is in agreement with the blackbody temperature $\sim 10^{5}~{\rm K}$ obtained from the X-ray luminosity.  

The bolometric wind and disc luminosities are shown in Table \ref{lbolt}. The wind bolometric luminosity, $L_w$, is given by equation (\ref{lwind}). 
The wind luminosity decrease with time, and the ratio of the disc to wind luminosity increases with time which implies that the contribution of disc luminosity to the spectrum increases with time. Due to significant standard errors in radiative efficiency, $\eta$, and $\Delta t$, the disc luminosity, and the luminosity ratio, have large ranges. The X-ray luminosity in the range $0.3 - 10~{\rm keV}$ is also shown. No significant X-ray emission was observed and an upper limit adopting a power-law spectrum with a photon index of $\Gamma = 2$ is $L_X < 3.3 \times 10^{41}~{\rm erg~s^{-1}}$ \citep{2017ApJ...842...29H}. The X-ray luminosity for the obtained parameters is $L_X \sim 4.5 \times 10^{42}~{\rm erg~s^{-1}}$, while the lower limit of obtained X-ray luminosity is $L_X \sim 1.18 \times 10^{39}~{\rm erg~s^{-1}}$ as can be seen from Table \ref{lbolt}. The minimum value of the derived X-ray luminosity implies insignificant X-ray emission which is in agreement with the observations.  

\begin{table}
\caption{The wind ($L_w$) and the disc ($L_d$) bolometric luminosity is shown for the obtained parameters. The X-ray luminosity in 0.3-10 keV from the disc is also shown. The super and subscripts show the maximum and minimum values. All the luminosities are in unit of $\rm{erg~s^{-1}}$.}
\label{lbolt}
\centering
\begin{tabular}{c|c|c|c|c|}
\hline
&&&&\\
$t~{\rm (days)}$ & $\log_{\rm 10}(L_{w})$ & $\log_{\rm 10}(L_{d})$ & $L_d/L_w$ & $\log_{\rm 10}(L_X)$\\
&&&&\\
\hline \\
&&&&\\
5  & $44.42_{-0.246}^{+0.197}$ & $43.98_{-1.227}^{+1.339}$ & $0.36_{0.014}^{13.04}$ & $42.65_{-3.584}^{+2.377}$\\
&&&&\\
11 & $44.37_{-0.251}^{+0.196}$ & $43.98_{-1.228}^{+1.335}$ & $0.403_{0.015}^{14.04}$ & $42.65_{-3.585}^{+2.371}$ \\
&&&&\\ 
14 & $44.36_{-0.253}^{+0.196}$ & $43.98_{-1.228}^{+1.329}$ & $0.42_{0.016}^{14.25}$ &  $42.66_{-3.586}^{+2.364}$ \\
&&&&\\
37 & $44.26_{-0.267}^{+0.195}$ & $43.98_{-1.231}^{+1.249}$ & $0.524_{0.020}^{14.71}$ & $42.66_{-3.592}^{+2.252}$ \\
&&&&\\
\hline
\end{tabular}
\end{table}

We assume a density profile $\rho(r) \propto r^{-2}$ for an outflowing wind with a constant velocity for both spectral line model using CLOUDY and disc-wind model for continuum spectrum. The disc-wind model for the continuum is simplified by assuming the Thomson opacity in the advective regime and a combination of Thomson and Kramers opacities in the diffusive regime (neglecting the details of scattering and absorption of photons in the medium). The model does not have the details of the atmosphere's elemental compositions which is crucial in deciding the opacity and the emission lines. The results from the spectral line model of CLOUDY are obtained by matching the observed emission lines of hydrogen and helium, whereas the results of the disc-wind model for continuum are based on fitting the continuum spectra from optical/UV bands assuming a blackbody emission. 
Thus, the comparison of two independent calculations is based on the atmospheric properties such as the density and temperature at the inner radius. 

The line luminosity fitting from CLOUDY results in the wind inner radius $r_{\rm in} = 7.079 \times 10^{14}~{\rm cm}$, whereas the disc-wind model fit to continuum spectrum results in $2.013 \times 10^{14}~{\rm cm}$. In the disc-wind model, the wind velocity is obtained using a $\chi^2$ model fit to the continuum, and in the CLOUDY modelling, we vary the free parameter $v_w$ close to its observed value derived from the FWHM of the HeII and H lines (see Fig. 15 in \citealp{2017ApJ...842...29H}). The constant wind velocity from CLOUDY is $1.3 \times 10^4~{\rm km~s^{-1}}$ whereas from the disc-wind model is $1.89 \times 10^4~{\rm km~s^{-1}}$. The results are of a similar order of magnitude and the model predicts a lower inner radius. We compare the atmosphere properties of the two independent calculations at the radius of CLOUDY. The helium abundance is super-solar to match the observed line luminosity of hydrogen and helium lines. We calculate the number density of the outflowing wind and the gas temperature from the model at the radius estimated by CLOUDY ($r_{\rm in} = 7.079 \times 10^{14}~{\rm cm}$), and compare the obtained values that are shown in Table \ref{dencomp}. The density at the considered radius decreases with time for both cases but the continuum spectrum fitting disc-wind model predicts higher values than the CLOUDY modelling.

\begin{table}
\caption{The number density of the gas and the temperature at the radius obtained $r_{\rm l} = 7.079 \times 10^{14}~{\rm cm}$, corresponding to the inner radius considered at CLOUDY. }
\label{dencomp}
\centering
\scalebox{0.8}{
\begin{tabular}{c|c|c|c|c|}
\hline
& \multicolumn{2}{|c|}{CLOUDY} & \multicolumn{2}{|c|}{Disc-wind} \\
\hline
&&&&\\
$t~{\rm (days)}$ & $\log_{\rm 10}(n~{\rm (cm^{-3})})$ & $\log_{\rm 10}(T(K))$ & $\log_{\rm 10}(n~{\rm (cm^{-3})})$ & $\log_{\rm 10}(T(K))$\\
&&&&\\
\hline \\
&&&&\\
5 & 10.01 & 4 & 10.90$_{-0.289}^{+0.214}$ & 4.77$_{-0.114}^{+0.064}$ \\
&&&&\\
11 & 10.00 & 4 & 10.74$_{-0.308}^{+0.206}$ & 4.73$_{-0.135}^{+0.063}$ \\
&&&&\\
14 & 9.95 & 4 & 10.67$_{-0.319}^{+0.204}$ & 4.71$_{-0.145}^{+0.062}$ \\
&&&&\\
37 & 9.91 & 4 & 10.28$_{-0.386}^{+0.203}$ & 4.61$_{-0.209}^{+0.074}$ \\
&&&&\\
\hline
\end{tabular}
}
\end{table}

In the CLOUDY calculations, we assume that the gas is irradiated by two temperatures of $10^5~{\rm K}$ and $10^4~{\rm K}$ \citep{2017ApJ...842...29H},  which correspond to X-ray emission from the disc and optical/UV emission from the outflow, respectively. The corresponding luminosity ratio is $L(10^5)/L(10^4) \simeq 0.05$, and implies that the luminosity from the disc is smaller than the luminosity from the wind. The luminosity ratio from the disc-wind modelling is given in Table \ref{lbolt}. The value considered for CLOUDY modelling is within the error limit of the values obtained using disc-wind modelling.

\section{Discussion}
\label{discussion}

The analysis of TDE spectra provides an understanding of the accretion processes, the radiative process in the medium, and the nature of the disrupted stars. The atmosphere plays a crucial role in the line luminosity and line ratios. The line luminosity in TDE spectra evolves with time which implies that the accretion dynamics and the atmosphere evolves. In this paper, we consider the TDE source with spectra dominated by the helium line over the hydrogen line. Earlier CLOUDY model fit to the PS1-10jh spectra \citep{2015MNRAS.454.2321S} showed that a super-solar composition is required whereas \citet{2016ApJ...827....3R} argued that the dominance of helium lines over hydrogen lines can be due to the high effective absorption optical depth. In this paper, we use TDE iPTF16axa whose spectrum has a helium line dominating over the hydrogen line and the line luminosity ratio evolves with time significantly. We model four different epochs using time-independent non-local Thermodynamic Equilibrium (NLTE) snapshot models in CLOUDY to reproduce the observed line luminosity and estimate the atmosphere properties and its elemental compositions. 
The Cloudy modelling of the spectrum on different days provides the density, temperature and optical depth in the medium along with the outflowing velocity. To interpret the CLOUDY solutions in TDE dynamics, we consider a steady disc-wind model where the disc is an advective slim disc with an accretion rate modulated by the outflow rate which evolves with time. The outflow evolves through radiative and diffusive regimes. \citet{2020ApJ...894....2P} showed that taking Thomson opacity in the diffusive regime predicts higher photosphere radius and thus, we also include the Kramers opacity due to absorption in the diffusive regime. We fit the disc-wind model with the spectrum continuum from optical/UV observations and estimated the accretion disc and outflow (atmosphere) properties. 

\citet{2017ApJ...842...29H} reported that no significant X-ray emissions were detected and the upper limit of 0.3-10 keV X-ray luminosity was $L_X < 3.3 \times 10^{41}~{\rm erg~s^{-1}}$. The X-ray emission can be either from the disc or the corona. The X-ray emission is from the disc if the X-ray spectrum is dominated by the blackbody emission and is from the corona if the X-ray spectrum shows power-law emission. \citet{2020MNRAS.497L...1W} has shown through the evolution of the spectral index ($\alpha_{\rm oX}$) from the TDE observations that the power-law X-ray emission dominates the spectrum when the luminosity ratio (bolometric to Eddington luminosity) is low, and the disc emission dominates the X-ray spectrum at high luminosity ratio. Here, we assume that the X-ray emission is from the disc and has neglected the corona. The inclusion of corona in the disc-wind model is complex and will require numerical modelling. 

In both CLOUDY and disc-wind modellings, it is assumed that the wind structure has a density profile given by $\rho \propto r^{-2}$ with the outflow velocity to be a constant. We obtain the photosphere radius and temperature of the emission that are comparable to the values obtained by \citet{2017ApJ...842...29H} using a single blackbody temperature to fit the continuum spectrum from optical/UV observations. Our disc-wind model fit results in a well-constrained stellar mass $M_{\star}$, wind inner radius $r_l$, and outflow velocity $v_w$, but is poorly constrained along efficiency $\eta$ and $\Delta t$ due to the high standard errors. These high errors cause significant uncertainty in the density, temperature, and luminosity. This is because the outflow rate $\dot{M}_{\rm out} = f_0(\dot{M}_{\rm fb}/M_{E})~\dot{M}_{\rm fb}$, where $\dot{M}_{E}$ depends on the efficiency $\eta$. 

In the wind-driven model presented in section \ref{dwmodel_theory}, the temperature at the launch radius is estimated by assuming equilibrium between the kinetic energy of the outflow and the internal energy dominated by the radiation. The photons are advected adiabatically until the trapping radius and then through the diffusion until the photosphere. The wind luminosity shown in Table \ref{lbolt} indicates the luminosity estimated using the kinetic energy of the outflow. Fig. \ref{mfnt} shows that the mass outflow rate $\dot{M}_{\rm out}$ increases with an increase in the $\dot{M}_{\rm fb}$ but the mass accretion rate become nearly constant at high $\dot{M}_{\rm fb}$. This implies that the disc luminosity is nearly constant when the total mass loss rate is high but the kinetic energy of the outflow increases resulting in an increase in the kinetic luminosity of the wind. \citet{2011MNRAS.410..359L} using an advective steady slim disc model to calculate the disc luminosity and a spherical wind model assuming the photons are advected adiabatically, showed that the wind kinetic luminosity exceeds the disc luminosity if the mass fallback rate is high and the disc luminosity is roughly constant when the wind luminosity is higher than the disc luminosity. Table \ref{lbolt} shows that the disc bolometric luminosity is nearly constant and the wind luminosity is higher than the disc bolometric luminosity, and the wind luminosity decreases with time. 

The NLTE radiative transfer modelling using CLOUDY provides the elemental abundance required to explain the observed luminosity, whereas the disc-wind modelling results in the stellar mass and the radiative efficiency in the accretion disc. The simultaneous analysis and determination of elemental composition and stellar mass are crucial to examine the type of disrupted star. The obtained stellar mass is $M_{\star} = 6.20 \pm 1.19 M_{\star}$. The helium abundance in the star is largely unconstrained, although an approximate constraint of helium abundance through measurements of low degree acoustic modes using Asteroseismology was proposed by \citet{2007MNRAS.375..861H}. In the case of TDEs, the observation is done after the star gets disrupted, and the CLOUDY fitting shows that a super-solar helium abundance is required to explain the observed line luminosities. We expect this for the obtained stellar mass. The higher helium abundance suggests an evolved red giant disrupted star.

 The emitted line photons at any radius in the atmosphere escape through a random walk scattering in the medium where they can be re-absorbed either in the line itself or by the continuum process. The photons absorbed have a probability of line scattering due to bound-bound transition or being destroyed via mechanisms such as de-excitation of atoms to a different line transition, photoionization, or excitation to another bound level. The scattering and absorption occur until the photons reach the thermalization radius (above which photons can scatter without absorption). The scattering and self-absorption increase with the optical depth which is a function of the density and temperature of the medium, and the photon frequency. Thus, the line luminosity may reduce if the optical depth is high. 

Table \ref{tab:table 3} shows the column density of various elements in the atmosphere and it is dominated by helium and hydrogen lines. In the CLOUDY simulations, we find that the mean line optical depth (from $r_l$ to $r_w$) of He II 4685.64 on day 5, 11, 14, and 37 are $3.99\times 10^{-2}$, $4.36 \times 10^{-2}$, $3.69 \times 10^{-2}$, and  $3.69 \times 10^{-2}$ respectively, whereas the mean optical depth for hydrogen is much higher. The large optical depths imply a higher degree of obscuration. \citet{2016ApJ...827....3R} showed through analytic formulation for a stationary atmosphere with photon diffusion that the outcome luminosity is $L \propto {\rm e}^{-\tau}B(\nu,~T)$, where $\tau$ is the optical depth corresponding to the emission line and $B(\nu,~T)$ is the Planck blackbody law with temperature $T$. By this, the high mean optical depth of hydrogen results in a larger obscuration and a lower luminosity. However, we would like to emphasize that the atmosphere in our case is non-stationary with a wind velocity of $v_w \sim {\rm few} \times 10^4~{\rm km~s^{-1}}$.  

In the presence of outflow, a highly dense reprocessing envelope can completely suppress the emission of H$\alpha$ line, but the envelope expands with time which impacts the optical depth and results in the time evolution of line luminosity. The CLOUDY calculation shows that the ratio $L({\rm He{\sc II}})/L({\rm H\alpha})$ varies from 1.82 on day 5 to 1.61 on day 37. Note that in the disc-wind model, the optical depth in the advective regime is given by $\tau \propto \dot{M}_{\rm out}$ for a constant velocity and is the same for both hydrogen and helium lines. However, the optical depth is decided by the radiative transfer mechanism, ionization temperature, and the element's abundance. Thus, the CLOUDY shows different mean optical depths for hydrogen and helium lines. 

The line spectra of iPTF16axa are symmetric as can be seen from Fig. 10 in \citet{2017ApJ...842...29H}. The outflow velocity in both the CLOUDY and disc-wind models is assumed to be constant and results in a spectrum where the lines are symmetric. The outflow velocity depends on the local radiative flux and the gravity in the disc and thus can vary with the disc radius. The radiative flux in a TDE disc evolves with time and thus, the outflow velocity may evolve with time. The velocity may also vary within the outflow and \citet{2018ApJ...855...54R} performed a radiative transfer calculation in spherical symmetry using Monte Carlo radiative transfer code and showed that the radially varying velocity results in an axisymmetry line emission. The determination of velocity and gas density as a function of space and time needs radiation hydrodynamic simulations. However, the constant velocity assumption provides atmosphere details that are consistent and in agreement with the observations. The CLOUDY modelling with constant velocity is able to explain the observed luminosities and the full width half maximum (FWHM) of hydrogen and helium lines (see Fig. \ref{fig:spectrum_fit}), whereas the disc-wind modelling with constant velocity is able to reproduce the continuum spectra (see Fig. \ref{lspecplt}).   

Finally, we note that in general, the TDE disc-wind model is non-steady and the outflow structure can be different from the spherical case with an outflow rate that may not be proportional to the mass fallback rate. The spherical outflow model provides a simplified geometry to understand wind dynamics. In our modelling for continuum spectrum fitting, we assume that the disc has an outflow at all radii and the mass outflow rate is the same at all radii at any given time. However, in a non-steady disc, the mass outflow rate varies with radius and time; this results in non-spherical outflows with varying wind velocity. Such modelling for optical/UV spectrum requires a time-dependent disc-wind model with radiative transfer. The simplified formulation presented in this paper is able to explain the observed lines and continuum. 

We have considered the TDE source iPTF16axa whose spectrum is dominated by helium and hydrogen lines while the other emission lines are significantly weak. However, TDEs  ASASSN-14li, PTF15af, and iPTF16fnl show strong nitrogen emission lines but weak or undetectable carbon lines \citep{2017ApJ...846..150Y}. Similarly, TDE AT2019qiz shows broad emission lines of hydrogen, He II (4686 \AA), the Bowen fluorescence lines of N III (4100 and 4640 \AA) and likely O III (3670 \AA) \citep{2020MNRAS.499..482N}. TDEs with strong nitrogen lines in their spectra are classified as N-rich TDE \citep{2019ApJ...887..218L}. In future, we plan to do more such TDE modelling using CLOUDY and include emission lines other than helium and hydrogen.  

\section{Summary}
\label{summary}

In this paper, we focus on studying the atmosphere properties and the elemental abundance required to explain the observed line luminosity of TDE iPTF6axa whose spectra are dominated by hydrogen and helium lines. We use the CLOUDY modelling on the line luminosities to obtain the density and temperature in the atmosphere and the abundance of elements (see Table \ref{tab:table 3}). We employ the accretion disc-wind model to conclude the evolution of atmospheric properties as obtained in CLOUDY from the perspective of TDE dynamics. We have performed independently the disc-wind model fit to the iPTF16axa spectra continuum and the CLOUDY modelling to the hydrogen and helium line luminosities.

We have assumed an outflow with constant velocity and density $\rho \propto r^{-2}$. In the CLOUDY modelling of the spectral lines, the irradiation of the gas in the outflow by two temperatures that are close to $10^5$ K and $10^4$ K is required. The outflow inner radius is $r_{\rm in} = 7.07 \times 10^{14}~{\rm cm}$ and the velocity is $v_w = 1.3 \times 10^4~{\rm km~s^{-1}}$. These values are in agreement with the values $r_{\rm in} = 2.01 \times 10^{14}~{\rm cm}$ and $v_w = 1.89 \times 10^4 ~{\rm km~s^{-1}}$ obtained using the disc-wind model to optical/UV continuum spectrum.

We vary He abundances keeping the abundances of other elements fixed at their solar abundances and the CLOUDY model predicts super solar abundance for He, which suggests that the disrupted star may be a red giant. The mean optical depth for helium is much smaller than the hydrogen and thus, the hydrogen line luminosity experience a larger obscuration than the helium line luminosity. 

The disc-wind model fit results in parameter values are $M_{\star} = 6.20 \pm 1.19 M_{\odot}$, $\log_{\rm 10}(\eta) = -1.22 \pm 1.327$, $r_l = (2.013 \pm 0.551) \times 10^{14}~{\rm cm}$, $v_w = 18999.4 \pm 1785.1 ~{\rm km~s^{-1}}$, and $\log_{\rm 10}(\Delta t ~{\rm (days)}) = 0.491 \pm 0.5$.

The single blackbody temperature from the disc using the parameter obtained from the disc-wind model fit is $T_d \sim 0.955 \times 10^5~{\rm K}$, which is in agreement with the temperature of $10^5~{\rm K}$ obtained from X-ray luminosity. The photosphere temperature in the wind is $\sim ~{\rm few} \times 10^4~{\rm K}$ (see Table \ref{rphtph}) which is in agreement with the observed value of $\sim 10^4~{\rm K}$ \citep{2017ApJ...842...29H}. 

The independent analyses for the atmosphere properties of the source iTPF16axa show comparable results that are in agreement with the observations. The CLOUDY photoionization modelling provides an abundance of various elements which is useful in predicting the stellar type of the disrupted star. The disc-wind model is required to estimate the stellar mass. The joint analysis using the disc-wind and CLOUDY modellings constrains the outflow properties such as density, temperature, and velocity, and is useful to probe the type of the disrupted star. 

\section{acknowledgments}
We thank Tiara Hung for providing data.
GS acknowledges WOS-A grant from the Department of Science and Technology (SR/WOS-A/PM-2/2021). 
MT has been supported by the Basic Science Research Program through the National Research Foundation of Korea (NRF) funded by the Ministry of Education (2016R1A5A1013277 and 2020R1A2C1007219). We thank the referee Prof. Andy Lawrence for his valuable comments and suggestions to improve our paper.

\section{Data Availability}
Simulations in this paper made use of the code CLOUDY (c17.02), which can be downloaded freely at https://www.nublado.org/. 
The model generated data are available on request.

\bibliographystyle{mnras}
\bibliography{ref} 


\bsp	
\label{lastpage}
\end{document}